\documentclass[twocolumn,prb,aps,epsf,showpacs,superscriptaddress]{revtex4-1}

\usepackage{graphicx}
\usepackage{dcolumn}
\usepackage{amsmath}
\usepackage{bm}

\newcommand{\figwidth}{0.9}

\newcommand{\figwidthtwo}{0.8}

\newcommand{\D}{\mathrm{d}}

\newcommand{\I}{\mathrm{i}}
\newcommand{\dwave}{$d$-wave}
\newcommand{\swave}{$s$-wave}
\newcommand{\tc}{$T_c$}
\newcommand{\ybco}[1]{YBa$_2$Cu$_3$O$_{#1}$}
\newcommand{\bscco}[1]{Bi$_2$Sr$_2$CaCu$_2$O$_{#1}$}

\newcommand{\tbcod}{Tl$_2$Ba$_2$CuO$_{6+\delta}$}

\begin{document}

\title{Microwave conductivity and superfluid density in strongly overdoped Tl$_{\bm 2}$Ba$_{\bm 2}$CuO$_{\bm{6 + \delta}}$}

\author{D.~Deepwell}
\affiliation{Department of Physics, Simon Fraser University, Burnaby, BC, V5A~1S6, Canada}
\author{D.~C.~Peets}
\affiliation{Department of Physics and Astronomy, University of British Columbia, Vancouver, BC, V6T~1Z1, Canada}
\affiliation{Department of Physics and Astronomy, Seoul National University, Seoul, 151-747, Korea}
\author{C.~J.~S.~Truncik}
\author{N.~C.~Murphy}
\author{M.~P.~Kennett}
\author{W.~A.~Huttema}
\affiliation{Department of Physics, Simon Fraser University, Burnaby, BC, V5A~1S6, Canada}
\author{Ruixing~Liang}
\author{D.~A.~Bonn}
\author{W.~N.~Hardy}
\affiliation{Department of Physics and Astronomy, University of British Columbia, Vancouver, BC, V6T~1Z1, Canada}
\affiliation{Canadian Institute for Advanced Research, Toronto, Ontario, MG5 1Z8, Canada}  
\author{D.~M.~Broun}
\affiliation{Department of Physics, Simon Fraser University, Burnaby, BC, V5A~1S6, Canada}

\begin{abstract}
We present measurements of the microwave surface impedance of the single-layer cuprate \tbcod, deep in the overdoped regime, with $T_c \approx 25$~K.  Measurements have been made using cavity perturbation of a dielectric resonator at 17 discrete frequencies ranging from 2.50 to 19.16~GHz, and at temperatures from 0.12 to 27.5~K.  From the surface impedance we obtain the microwave conductivity, penetration depth and superfluid density.  The superfluid density displays a strong linear temperature dependence from 2 to 14~K, indicative of line nodes in the energy gap.  The microwave data are compared with theoretical predictions for a $d$-wave superconductor with point-like impurities, with the conclusion that disorder in \tbcod\ acts predominantly in the weak-to-intermediate-strength scattering regime, and that small-angle scattering is important.
\end{abstract}
 
 
\maketitle{} 
  
\section{Introduction}

An important obstacle in the problem of cuprate high-temperature superconductivity is our incomplete understanding of the normal state from which superconductivity appears.\cite{Orenstein:2000tw,Bonn:2006ue}  At optimal doping, the normal state is a strange metal,\cite{VARMA:1989p468} with linear-in-temperature resistivity,\cite{Gurvitch:1987ck} strong inelastic scattering\cite{Valla:1999p641} and uncertain electronic structure.\cite{Daou:2008iq}  On the underdoped side, the normal phase is characterized by a prominent pseudogap\cite{Timusk:1999p422} and a reconstructed Fermi surface.\cite{Chang:2010ic,DoironLeyraud:2013kg}  Only in strongly overdoped materials has a clear understanding of the normal state emerged, and this has been largely due to work on a single material: the single-layer cuprate \tbcod.\cite{Peets:2007gc}  The picture that has developed in this material is of a more-or-less conventional Fermi liquid, with a single, large Fermi surface.  

On the approach to the overdoped side, there is evidence for a significant change in the electronic structure around a doping level of $p = 0.21$ holes per planar copper.  The behaviour of the resistivity is suggestive of fluctuations around a quantum critical point, with an unconventional linear term that is strongest at dopings around $p \approx 0.2$.\cite{Daou:2008iq,Cooper:2009dg}  Spectroscopic studies reveal key differences between optimally doped $(p \le 0.2)$ and overdoped $(p > 0.2)$ material: there is a significant change in the doping-dependence of the low-energy spectral weight associated with Zhang--Rice singlets;\cite{Peets:2009iv,Schneider:2005ky} and angle-resolved photoemission (ARPES) reveals a reversal in the nodal/antinodal dichotomy --- in contrast to the situation in underdoped and optimally doped material, the quasiparticle line shape is sharp in the antinodal regions, and broad along the nodal directions.\cite{Peets:2007gc}  Carrying out experiments that can connect the optimal and overdoped regimes is now of primary importance.

Evidence for Fermi-liquid-like behaviour in strongly overdoped \tbcod\ was first obtained from magnetotransport studies, where a DC resistivity of the form $\rho(T) \approx \rho_0 + A T^{1.75}$ was reported.\cite{MACKENZIE:1993p197}  Hall-effect measurements then revealed the existence of a large Fermi surface,\cite{Mackenzie:1996p199} which has subsequently been confirmed by angle-dependent magnetoresistance oscillations (AMRO)\cite{Hussey:2003ck,AbdelJawad:2006df,Analytis:2007ja,Kennett:2007hx} and ARPES.\cite{Plate:2005p2658}  More recently, quantum oscillations have been observed in \tbcod.\cite{Vignolle:2008p1694,Bangura:2010p1675} In addition to giving detailed information on the structure of the Fermi surface, the quantum oscillation measurements provide a striking confirmation of the Fermi-liquid picture, with the temperature damping of the quantum-oscillation amplitude closely following the Lifshitz--Kosevich form expected for a Fermi liquid.\cite{Shoenberg:1984wo}

Despite the conceptual simplicity of the overdoped side of the cuprate phase diagram, relatively few experiments have been carried out in the superconducting state of strongly overdoped \tbcod.   This is in part due to rather severe materials synthesis challenges arising from the volatility and toxicity of thallium-based compounds, which limit the size and availability of \tbcod\ crystals.   One notable group of experiments has addressed the nature of the superconducting field--temperature phase diagram in strongly overdoped \tbcod.  Magnetotransport measurements initially revealed a resistive upper critical field, $B_\rho(T)$, with strong upwards curvature across the entire temperature range.\cite{MACKENZIE:1993p197}   Specific heat measurements\cite{Carrington:1996df} then demonstrated that superconducting order persists well beyond the $B_\rho(T)$ boundary, raising the possibility that $B_\rho(T)$ demarcates the melting of the vortex lattice.  However, electrical transport immediately beyond $B_\rho(T)$ is indistinguishable from that in the normal state,\cite{MACKENZIE:1993p197} and this is difficult to reconcile with the presence of a conventional vortex liquid.  Magnetization measurements\cite{Bergemann:1998gl} confirm this, ruling out a simple, London-like vortex liquid and instead revealing a linear diamagnetic response that persists to at least 10~T.    In Ref.~\onlinecite{Geshkenbein:1998p3010}, Geshkenbein and co-workers point out that this behaviour can be understood if there are simultaneously ``cold-spots'' in the quasiparticle scattering rate, making the vortex viscosity anomalously small, and microscopic inhomogeneities, several hundred angstroms in size, at which the local $T_c$ is much higher than in the bulk.  Microscopic inhomogeneity\cite{2001SSCom.120..347U} has also been considered as an explanation for the decrease in superfluid density in strongly overdoped \tbcod\ observed in muon spin-relaxation ($\mu$SR) measurements.\cite{Uemura:1993gs}  The microwave data we present below on strongly overdoped \tbcod\ can be used to test and constrain these ideas.

On the issue of pairing symmetry in \tbcod, most is known from experiments that have been carried out near optimal doping, with all evidence pointing to a $d$-wave pairing state. Penetration depth measurements, in near-optimally doped ($T_c =78$~K) \tbcod, revealed a strong linear temperature dependence of the superfluid density, implying the presence of line nodes in the energy gap.\cite{Broun:1997p387} These were followed by scanning SQUID measurements of half-integer flux quanta in \tbcod\ grown on a tetra-crystal substrate, which provided a phase-sensitive determination of pure $d$-wave symmetry.\cite{Tsuei:1998fk}  Only in the case of thermal conductivity have superconducting-state experiments been pushed into the strongly overdoped regime.\cite{Proust:P2lqZi4f,Hawthorn:2007hq}  These reveal another key signature of $d$-wave superconductivity --- universal heat conduction due to zero-energy $d$-wave quasiparticles --- and show that the superconducting gap energy, $\Delta$, scales with $T_c$ in this part of the phase diagram.

In summary, the situation is that most work on pairing symmetry and quasiparticle charge dynamics in the cuprates has been carried out at low-to-optimal dopings, while the Fermiology of the normal state is best understood at high doping.  The preponderance of Fermi surface data for samples with $T_c \approx 25$~to~30~K makes this particular doping level an opportune connection point.  We have therefore carried out a study of the microwave conductivity and penetration depth of a single crystal of \tbcod\ with $T_c \approx 25$~K, extending the range in which electrodynamic data exist deep into the overdoped regime.   Clear signatures of $d$-wave superconductivity are observed, notably a linear temperature dependence of the superfluid density.  The availability of detailed normal-state data allows a quantitative comparison to be made with theoretical calculations for a $d$-wave superconductor.  This has been carried out using results from self-consistent $t$-matrix approximation (SCTMA)\cite{NAM:1967p640,PETHICK:1986p562,HIRSCHFELD:1988p611,PROHAMMER:1991p557,Schachinger:2003p635,HIRSCHFELD:1993tf,HIRSCHFELD:1993p567,BORKOWSKI:1994p647,Joynt:1997p601,Hussey:2002p649,Nunner:2005p654,Balatsky:2006p607} theory for point-like disorder,
and a consistent picture of the charge dynamics emerges.

\section{Experimental methods}
 
\subsection{Tl$_\mathbf{2}$Ba$_\mathbf{2}$CuO$_\mathbf{6 + \delta}$ samples}

Single crystals of the single-layer thallium cuprate have nominal composition \tbcod: throughout this paper we will refer to the material in this manner.  However, the actual material grows with the approximate chemical composition Tl$_{1.85}$Ba$_2$Cu$_{1.15}$O$_{6 + \delta}$ --- it is stabilized by an excess of Cu, which substitutes for Tl in the TlO$_2$ layers.\cite{Shimakawa:1993jm}  Oxygen interstitials, which are important for tuning hole doping, are also located in the TlO$_2$ layers.   The impurity potentials corresponding to this type of disorder are softened and weakened by the screening effect of the BaO layers between the TlO$_2$ layers and the CuO$_2$ planes: we would therefore expect the disorder to act in the weak-scattering (Born) limit; to act as a source of small-angle scattering; and to correspond to a high density of scatterers.\cite{Abrahams:2000hr,Nunner:2005p654}

For this study, single crystals of \tbcod\ were grown by a time-varying encapsulation scheme\cite{Peets:2007p2147,Peets:2010p2131} using a copper-oxide-rich self-flux method in Al$_2$O$_3$ crucibles.  A sample with as-grown surfaces was used for the microwave measurements reported here, with in-plane area 0.39~mm$^2$ and thickness 11~$\mu$m.  It was annealed in pure flowing oxygen at 500~$^\circ$C for 10~days, to produce an overdoped sample with $T_c \approx 25$~K.  The sample was embedded in \tbcod\ powder and placed in a small quartz test-tube during the oxygenation anneal, which was quenched in an ice-water bath at the end of the annealing run to ensure good oxygen homogeneity. 

\subsection{Surface impedance measurements}  
\label{Sec:surface_impedance}

Microwave surface impedance, $Z_s = R_s + \I X_s$, was measured using cavity perturbation\cite{1946Natur.158..234P,Altshuler:1963wq,Klein:1993p1129,Donovan:1993p1114,DRESSEL:1993p341,Huttema:2006p344,Bonn:2007hl} of a rutile (TiO$_2$) resonator.\cite{Huttema:2006p344}  The resonator was a right, circular cylinder of diameter 10~mm and height 10.2~mm, with a 3~mm hole bored through the resonator along the cylinder axis.  The cylinder axis was aligned with the $c$-axis of the TiO$_2$, allowing the axially symmetric microwave modes used in our measurements to be modelled using a scalar dielectric constant.  Both the growth of the single-crystal TiO$_2$, and the fabrication of the resonator, were carried out by ELAN~Ltd., Russia.  The resonator was sandwiched between 2 sapphire plates, top and bottom, in the centre of a copper enclosure, taking care to make sure the overall structure retained reflection symmetry in the mid-plane of the resonator. Data were taken at a set of 17 discrete frequencies ranging from $\omega/2 \pi = 2.50$ to 19.16~GHz, as shown in Figs.~\ref{fig:Rs} and \ref{fig:Xs}. Measurements were made using the TE$_{0np}$ modes of the resonator ($p = 1,3$). These microwave modes are best pictured as standing waves of a cylindrical waveguide, with the mode number $p$ giving the number of half wavelengths along the $z$-direction.  By working with $p= 1$ and $p=3$ families of resonant modes, we guarantee that the \mbox{$z$-component} of microwave magnetic field has a local maximum at the centre of the resonator, at which $H_\mathrm{rf}$ points along the cylinder axis.  The rest of the frequency variation is obtained by scanning the radial mode number $n$.   Resonator temperature is kept fixed at $T \approx 1.5$~K during the measurements.

The platelet \tbcod\ sample was mounted at the end of a high-purity silicon hot-finger using vacuum grease, and then introduced into the resonator through a hole bored along the axis of the rutile cylinder.  All data sets were taken with $H_\mathrm{rf}$ oriented parallel to the crystal $c$-axis, to induce in-plane screening currents. This leads to large demagnetizing effects in our geometry, so we were careful to keep microwave power levels low enough to remain in the linear-response regime.  For separate reasons, we hold $H_\mathrm{rf}$ constant during temperature sweeps, to prevent small nonlinearities associated with the rutile resonator from contributing to the apparent temperature dependence of the sample's surface impedance.  This is in keeping with the golden rule of precision cavity perturbation measurement: to vary one, and only one, property of the resonant system at a time.

The microwave resonator system was mounted in an Oxford Instruments MX40 dilution refrigerator, as shown in Ref.~\onlinecite{Truncik:2013hr}, with the hot-finger technique allowing the temperature of the sample to be varied between 0.12 and 30~K while the resonator was kept at a fixed temperature of 1.5~K.  As a hot-finger material, silicon has advantages over sapphire for low-temperature work: signal contamination from paramagnetic effects is negligible, in contrast to prominent Curie-like behaviour in sapphire below 1~K.  There is a trade-off at high temperatures, however: thermal excitation of free carriers in the silicon became observable above 28~K, imposing a practical upper bound on the temperature range of the experiment. 

The temperature dependence of the surface impedance was inferred using the cavity perturbation relation\cite{Altshuler:1963wq,Klein:1993p1129,Huttema:2006p344,Bonn:2007hl}
\begin{equation}
\Delta R_s(T) + \I \Delta X_s(T) = \mathcal{G} \left(\tfrac{1}{2} \Delta f_B(T) - \I \Delta f_0(T) \right)\;.
\end{equation}
Here, $f_B$ is the resonant bandwidth of the TE$_{0np}$ mode of interest, $f_0$ is the corresponding resonant frequency, and temperature-dependent changes are with respect to the values of the variables at a reference temperature, $T_\mathrm{ref}$.  $\mathcal{G}$ is an empirically determined scale factor, different for each mode.  The resonator technique is accurate and fast, allowing measurements to be made in sequence at the 17 different frequencies but, on its own, does not always allow the absolute magnitude of $R_s$ and $X_s$ to be reliably determined.  In some cases, for lower-order microwave modes, a reasonably accurate measurement of absolute surface resistance can be made by subtracting the empty-resonator $f_B$ from the value with the sample in place.  However, experience has shown that this technique breaks down for the higher-order microwave modes.  To circumvent this problem, we have developed a bolometric technique that infers the absolute surface resistance from the temperature rise of the sample in the presence of a microwave magnetic field of known intensity, in a similar spirit to Ref.~\onlinecite{Turner:2004p332}.  The bolometric measurement is slow compared to a cavity perturbation measurement, but only needs to be carried out once for each frequency.  The absolute surface reactance is obtained using an \mbox{$R_s$:$X_s$-matching} technique to determine the zero-temperature penetration depth, $\lambda_0$.  That is, we find the temperature-independent offset, $X_{s0} = \omega \mu_0 \lambda_0$, which, when added to $\Delta X_s(T)$, imposes the Hagen--Rubens criterion that $R_s$ and $X_s$ be equal in the normal state.\cite{Klein:1993p1129}
The in-plane normal-state resistivity, $\rho_\mathrm{dc}(T=27.5~\mathrm{K}) = 10.2~\mu\Omega$cm, obtained in a separate measurement on the same sample, was used to determine the resonator constant $\mathcal{G}$ at each frequency by imposing a normalization condition on the real part of the microwave conductivity: $\sigma_1(T = 27.5~\mathrm{K}) = 1/\rho_\mathrm{dc}(T = 27.5~\mathrm{K})$.  Inaccuracy in the determination of the dc resistivity is the principal source of uncertainty in our measurement, and acts as a scale error for the overall data set.  We estimate the uncertainty in resistivity to be 10\%, leading to 5\% errors in surface impedance and penetration depth; and 10\% errors in conductivity and superfluid density.

With a thickness of 11~$\mu$m, the \tbcod\ sample is thin enough that we need to consider finite-size effects above \tc;  the normal state skin depth, $\delta = \sqrt{\rho_\mathrm{dc}/2 \omega \mu_0}$, is 1.6~$\mu$m at 2.50~GHz, 1.2~$\mu$m at 4.28~GHz, decreasing to 0.6~$\mu$m at the highest frequency, 19.16~GHz.   To correct the surface impedance data for finite-size, we use the following formula for effective surface impedance, $Z_s^\mathrm{eff}$, in a sample of thickness $t$:\cite{Waldram:1999p2205}
\begin{equation}
Z_s^\mathrm{eff} = Z_s^0 \coth\left(\I \omega \mu_0 t/2 Z_s^0\right)\;.
\label{Eq:finite_size}
\end{equation}
Here $Z_s^0$ is the bare surface impedance of a thick sample, and the $\coth(x)$ form of the finite-size expression is the one appropriate to the high-demagnetizing geometry used in our measurements.  Equation~\ref{Eq:finite_size} is appropriate when the 2D skin depth, $\delta^2/t$, is much smaller than the in-plane dimension of the sample, and reverts to the bare surface impedance $Z_s^0$ when $t \gg \delta$.  Our crystal is never very far into the finite-size limit, at any temperature or frequency, so the corrections are weak and all conditions are satisfied.  Nevertheless, we have carried out this correction on all the data.  This is important because the normal-state surface impedance is used both for calibration purposes and for the determination of $\lambda_0$.

\section{Results}
 
\subsection{Surface impedance}

\begin{figure}[t]
\centering
\includegraphics[width = \figwidth \columnwidth]{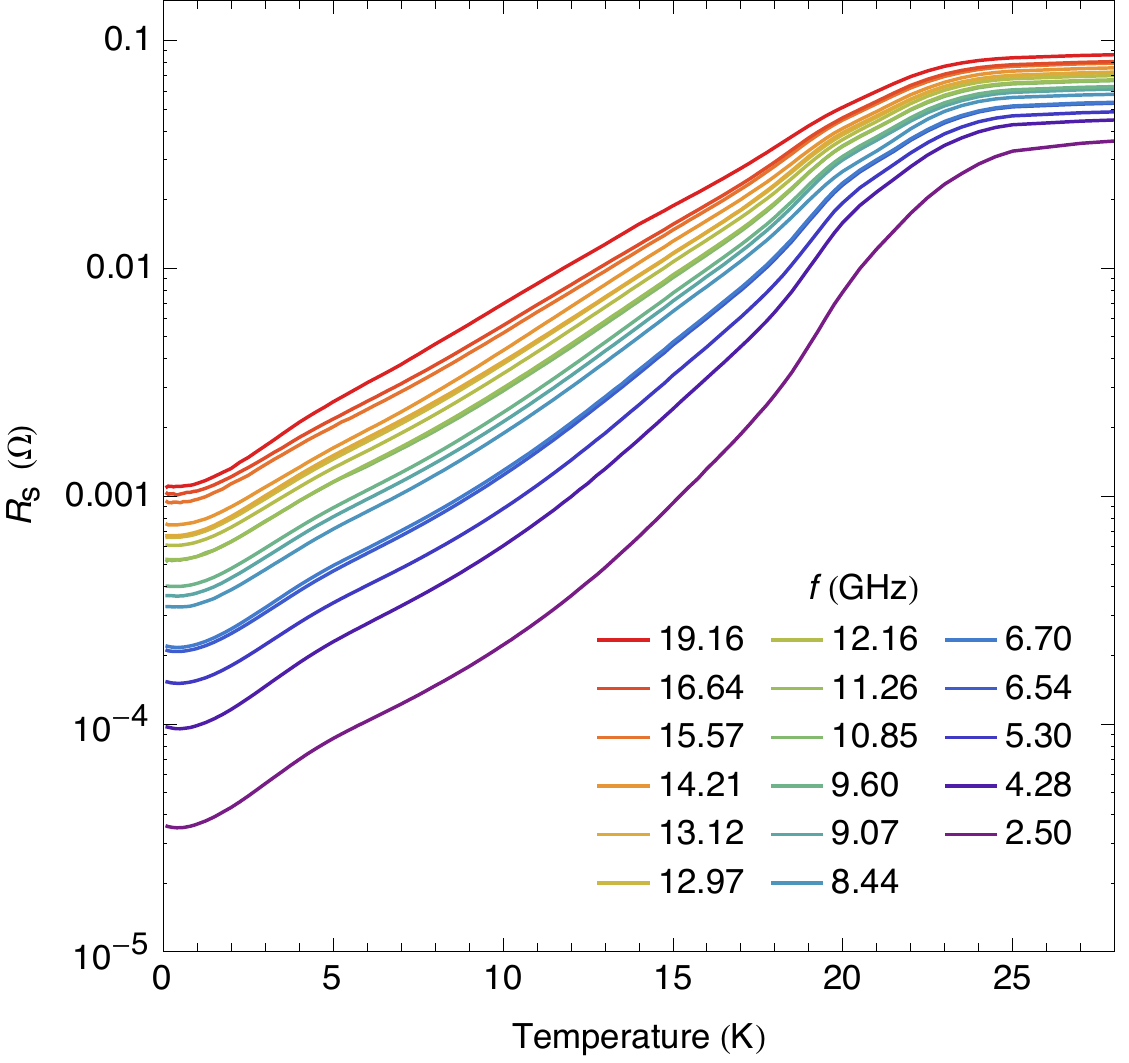}
\caption{(Color online) Surface resistance of \tbcod\ at a set of 17 frequencies ranging from 2.50 to 19.16~GHz, as indicated in the legend.  $R_s(T)$ decreases monotonically with temperature, with a rounded transition near \tc. The frequency dependence of $R_s$ becomes stronger at low temperatures, as the material transitions from the $R_s \propto \sqrt{\omega}$ behaviour of a metal to the $R_s \propto \omega^2$ regime of a superconductor.} 
\label{fig:Rs}
\end{figure}

The surface resistance, $R_s(T)$, is plotted in Fig.~\ref{fig:Rs} for the 17 frequencies.  $R_s$ shows relatively little temperature dependence above \tc, then begins a rounded transition into the superconducting state around 25~K.  Although fluctuation paraconductivity can be large in the cuprates,\cite{LANG:1994p246,Leridon:2007p240} it should be less prominent in overdoped materials: as we will later argue, the rounding in $R_s(T)$ is most likely associated with \tc\ variation in the sample, driven by chemical inhomogeneity. Below \tc, surface resistance decreases monotonically with temperature.  Several weak, broad features are visible in the $R_s(T)$ traces --- these appear more prominently in the real part of the microwave conductivity.  At low temperatures, there is substantial residual surface resistance. Below 5~GHz there are indications of small upturns in $R_s(T)$, suggestive of a small amount of paramagnetic absorption, perhaps from residual flux on the sample surface.  Above \tc, the plotted $R_s$ necessarily follows a $\sqrt{\omega}$ frequency dependence as a consequence of our normalization procedure, for which we assume $R_s(\omega) \approx \sqrt{\omega \mu_0 \rho_\mathrm{dc}/2}$.  Well below \tc, $R_s(\omega) \sim \omega^2$, the usual behaviour for a superconductor in the limit where $\omega$ is less than the quasiparticle relaxation rate.

\begin{figure}[t]
\centering
\includegraphics[width = \figwidth \columnwidth]{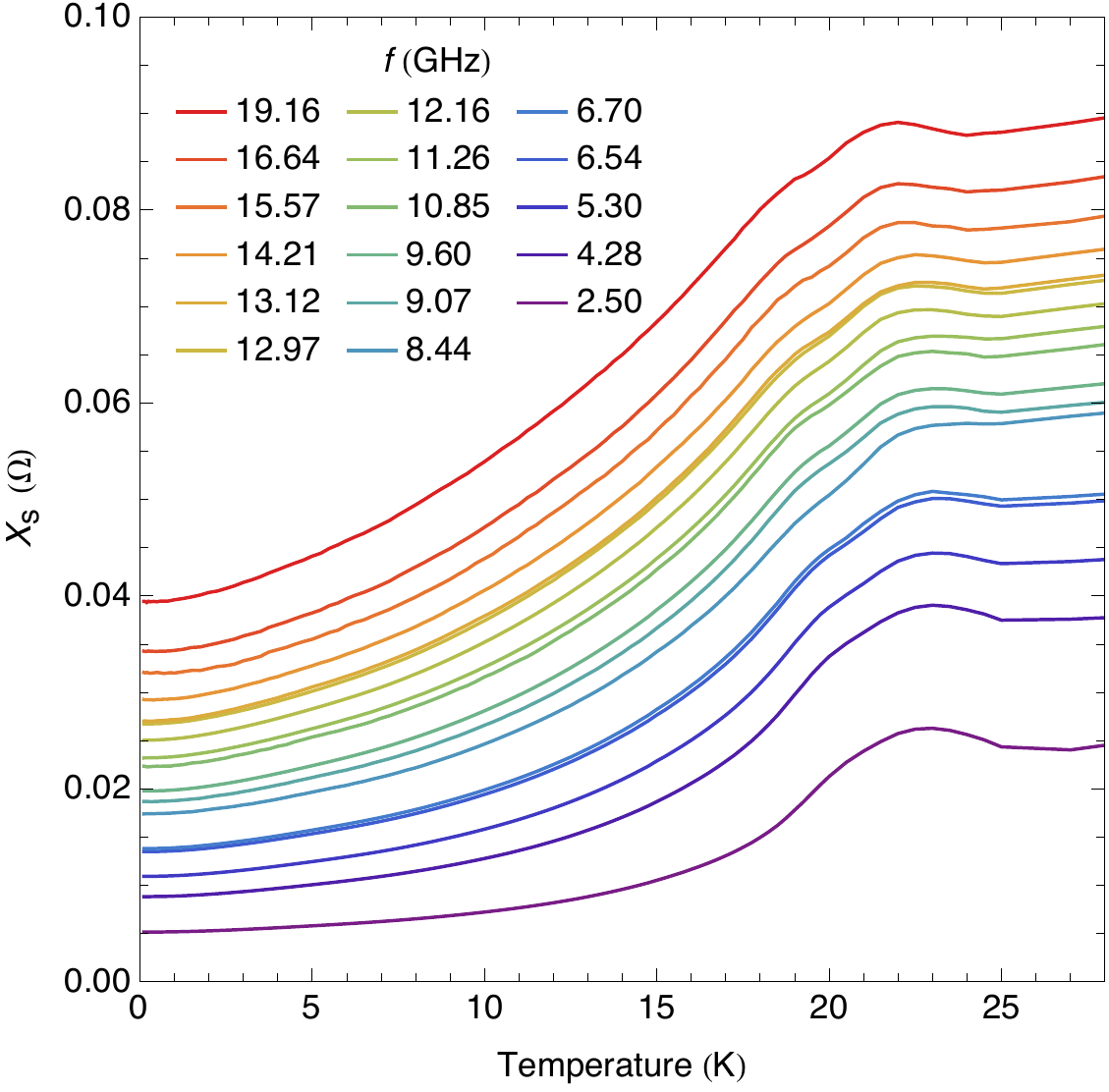}
\caption{(Color online) Surface reactance of \tbcod\ at the same set of frequencies as in Fig.~\ref{fig:Rs}, from 2.50~GHz (bottom) to 19.16~GHz (top).  At the onset of superconductivity, there is a small rise in $X_s(T)$ due to the kinetic inductance of the superconducting electrons.  As the electrons condense further, the penetration depth shrinks, causing $X_s(T)$ to decrease at lower temperatures.} 
\label{fig:Xs}
\end{figure}  

Surface reactance, $X_s(T)$, is plotted in Fig.~\ref{fig:Xs}.  As described in Sec.~\ref{Sec:surface_impedance}, temperature-dependent changes in surface reactance have been obtained by measuring the frequency shift of the dielectric resonator in the 17 TE$_{0np}$ resonant modes used in this experiment.  The most prominent feature in the temperature dependence of $X_s$ is a small peak in $X_s(T)$ immediately below \tc.  This is due to the onset of a kinetic-inductance contribution to the reactance, and is often observed at the superconducting transition. 
The absolute surface reactance has been obtained using the normal-state surface-impedance matching technique.  The zero-temperature surface reactance, $X_s(0) = \omega \mu_0 \lambda_0$, that gives the best match between $R_s(T)$ and $X_s(T)$ corresponds to a zero-temperature penetration depth $\lambda_0 = 2610\pm5$\%~\AA.  

\subsection{Microwave conductivity}

\begin{figure}[t]
\centering
\includegraphics[width = \figwidth \columnwidth]{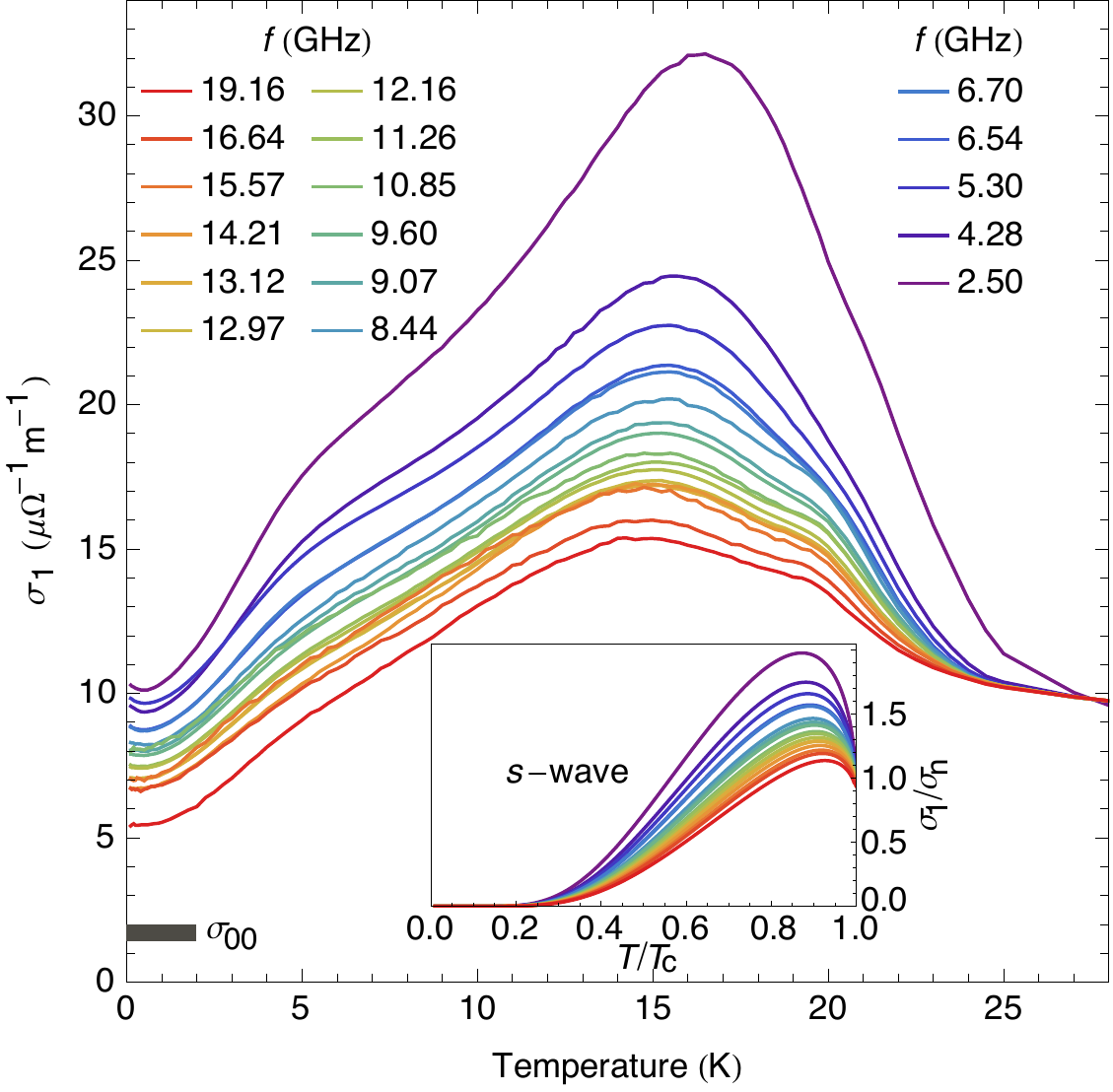}
\caption{(Color online) Real part of the microwave conductivity, $\sigma_1(T)$, at fixed frequencies from 2.50~GHz (top) to 19.16~GHz (bottom).  $\sigma_1(T)$ rises substantially on cooling through \tc, before peaking at 15 to 16~K, then decreasing to a residual low temperature value comparable to that in the normal state.  $\sigma_{00}$ denotes the bare universal conductivity expected for a $d$-wave conductor.
Inset: the microwave conductivity expected for an $s$-wave superconductor, calculated using Mattis--Bardeen theory\cite{MATTIS:1958p326} for the same set of reduced frequencies used in the \tbcod\ experiment.  In contrast to what is seen in \tbcod, the initial rise in $\sigma_1(T)$ on cooling is almost vertical, before peaking then becoming exponentially small at low temperature.} 
\label{fig:sigma1}
\end{figure}

The complex microwave conductivity, $\sigma = \sigma_1 - \I \sigma_2$, is obtained from the surface impedance using the local electrodynamic relation
\begin{equation}
Z_s = R_s + \I X_s = \sqrt{\frac{\I \omega \mu_0}{\sigma_1 - \I \sigma_2}}\;.
\label{Eq:local_electrodynamics}
\end{equation}
Equation~\ref{Eq:local_electrodynamics} is valid when the electromagnetic fields vary slowly over electronic lengths scales: i.e., when skin depth, $\delta$, and penetration depth $\lambda$ are much greater than coherence length, $\xi_0$, and electronic mean free path, $\ell$.  These constraints should be well satisfied in overdoped \tbcod.  The real part of the microwave conductivity, $\sigma_1(T)$, is plotted in Fig.~\ref{fig:sigma1} for the 17 frequencies.  The measured  $\sigma_1(T)$ traces evolve smoothly with frequency, giving us confidence in the use of the higher order TE$_{0np}$ modes for microwave spectroscopy.  The general behaviour is qualitatively similar at all frequencies: on cooling through \tc, there is an initial rise in $\sigma_1(T)$, followed by a peak in the range of 15 to 16~K and then a monotonic decrease to low temperatures. Below 3~K, $\sigma_1(T)$ follows a quadratic temperature dependence.  

\begin{figure}[t]
\centering
\includegraphics[width = \figwidth \columnwidth]{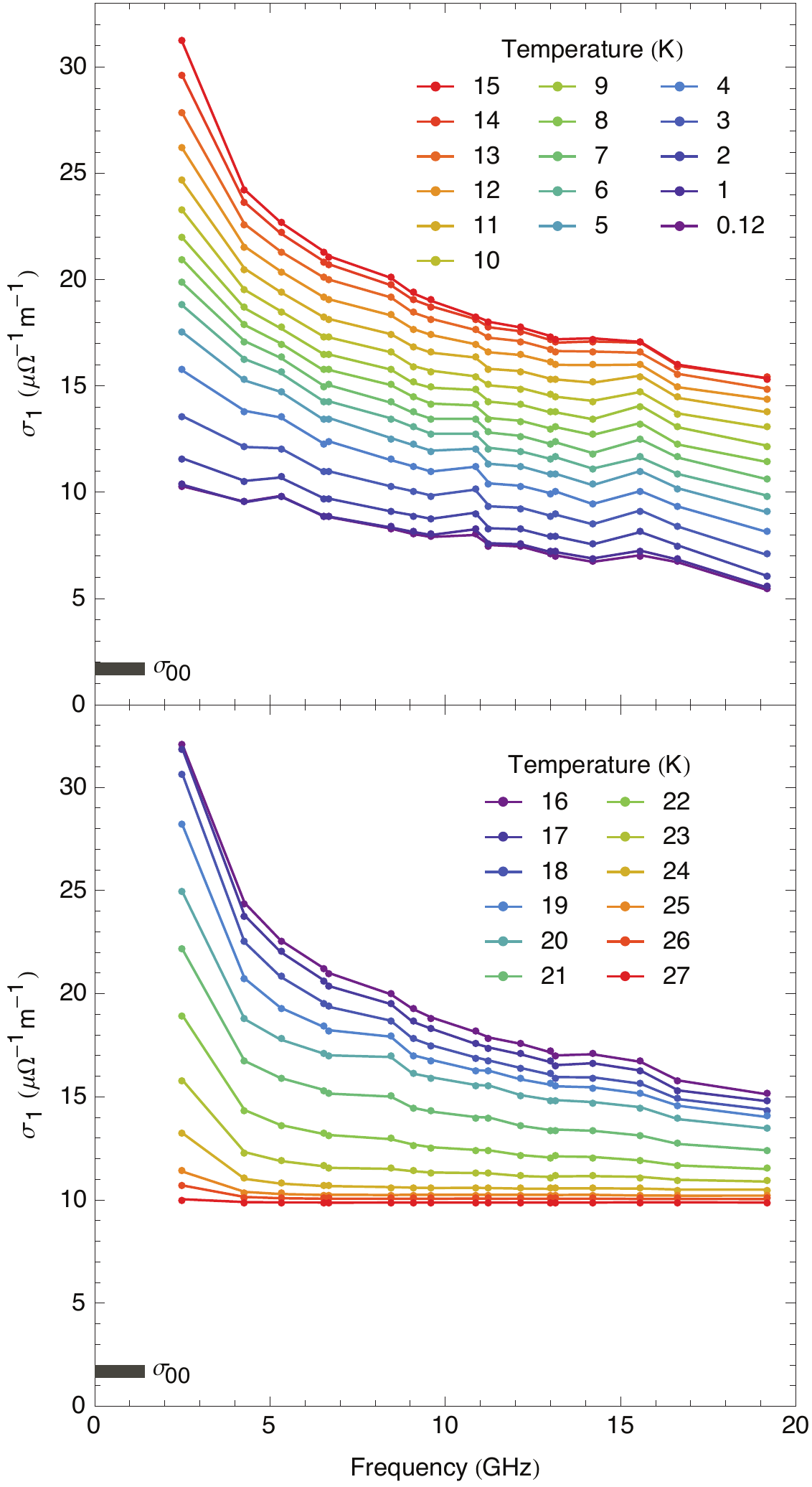}
\caption{(Color online) Microwave conductivity spectra, $\sigma_1(\omega)$, plotted at discrete temperatures. The strong variation of $\sigma_1(\omega)$ in the low GHz range indicates charge excitations relaxing on microwave timescales.  $\sigma_1$ decreases, and its frequency dependence weakens, on cooling to the lowest temperatures. The strong frequency dependence observed below \tc\ weakens then disappears on warming into the normal state.  (The perfectly flat behaviour at 27~K reflects the normalization condition, $\sigma_1(T = 27.5~\mathrm{K}) = 1/\rho_\mathrm{dc}$, used to determine the resonator constant at each frequency.) $\sigma_{00}$ denotes the bare universal conductivity expected for a $d$-wave conductor.} 
\label{fig:spectra}
\end{figure} 

The shape of  $\sigma_1(T)$ is very different from that of a BCS $s$-wave superconductor: for comparison, $\sigma_1(\omega,T)$ is plotted in the inset of Fig.~\ref{fig:sigma1} using Mattis--Bardeen theory\cite{MATTIS:1958p326} and displays an almost vertical rise immediately below \tc.  In other cuprate materials for which the microwave conductivity has been studied, most notably the \ybco{y} family, the microwave conductivity also displays a peaked structure as a function of temperature.\cite{Bonn:1992fx,Lee:1996p382,Broun:1997p387,Hosseini:1999p383,Ozcan:2006bm,Harris:2006p388}  In those systems, the behaviour of $\sigma_1(T)$ can be attributed to quasiparticle dynamics: the initial rise in $\sigma_1(T)$ on cooling through \tc\ is caused by a rapid decrease in \emph{inelastic scattering}; $\sigma_1(T)$ peaks where the quasiparticle relaxation rate reaches an elastic disorder limit; and then $\sigma_1(T)$ decreases at lower temperatures as the remaining quasiparticles condense.   The situation in overdoped \tbcod, although superficially similar, emerges as being quite different on closer inspection.  To begin with, the normal-state scattering is predominantly elastic (disorder limited) by the time material reaches \tc: there may be some additional decrease in quasiparticle scattering on entering the superconducting state, but this is more likely a result of reduced phase space for recoil than a collapse in inelastic scattering.  

Further clues as to the nature of the charge dynamics come from examining the conductivity spectrum, $\sigma_1(\omega)$, which we are able to plot in detail using the multiple-frequency resonator data.  This is done in Fig.~\ref{fig:spectra}.  At most temperatures, the $\sigma_1(\omega)$ spectra show strong frequency dependence in the microwave range --- this is a clear indication of long-lived charge excitations, relaxing on timescales of the order of 40~picoseconds.  However, for the lowest temperatures, we see that the frequency dependence of $\sigma_1$ is quite weak --- this is something of a paradox, since the quasiparticle relaxation rate should be lowest at low $T$.  This raises the question as to what  is responsible for the strong variation of $\sigma(\omega)$ in the microwave range --- our best hypothesis is that we are observing charge transport by short-lived superconducting currents, resulting from a combination of intrinsic dynamical inhomogeneities due to fluctuations, and static inhomogeneities arising from a spatial variation of \tc.  In fact, the microwave frequency scale, $f_\mathrm{fl}$, allows us to estimate the length scale over which these fluctuations occur: the fluctuation lifetime is $\tau_\mathrm{fl} = 1/2 \pi f_\mathrm{fl}$; the fluctuation length scale should be $\ell_\mathrm{fl} \approx v_F \tau_\mathrm{fl}$.  Using a Fermi velocity $v_F = 1.7 \times 10^5$~m/s (Ref.~\onlinecite{Bangura:2010p1675}), a fluctuation frequency scale of 4~GHz corresponds to a length scale of 7~$\mu$m.  This is far too long to be the quasiparticle mean free path, but it is quite feasible that superconducting inhomogeneities (dynamic or static) occur on this scale.  While this points to an increased role for inhomogeneities in overdoped \tbcod\ compared to optimally doped cuprates, as has been suggested in the context of superfluid density suppression,\cite{Uemura:1993gs,2001SSCom.120..347U} it is important to note that the narrow peaks in $\sigma(\omega)$ contain a negligible fraction of the total conductivity spectral weight --- we will later make an estimate, based on a spectral weight argument, that places the actual quasiparticle scattering rate in the range of 500~GHz.

\subsection{Superfluid density}

\begin{figure}[t]
\centering
\includegraphics[width = \figwidth \columnwidth]{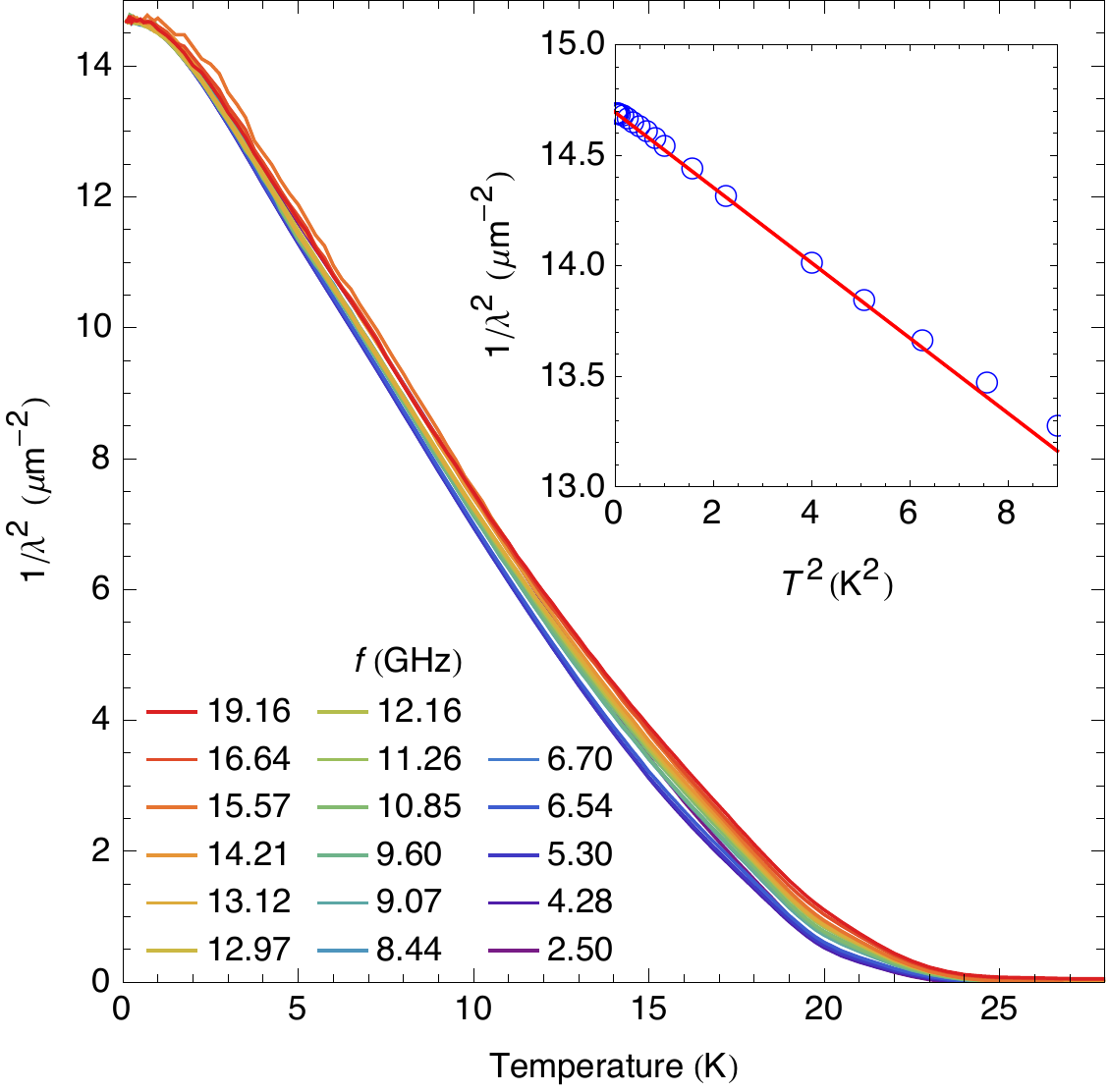}
\caption{(Color online) Frequency dependent superfluid density, $1/\lambda^2(T)$, from 2.50~GHz (bottom) to 19.16~GHz (top).  Over most of the temperature range, $1/\lambda^2(T)$ shows a strong linear temperature dependence, the expected behaviour of a $d$-wave superconductor with line nodes in the energy gap.  $1/\lambda^2(T)$ develops upward curvature on the approach to \tc: this, and the presence of fine structure in $1/\lambda^2(T)$, suggests some inhomogeneity of \tc\ across the sample.   At all temperatures, $1/\lambda^2$ is an increasing function of frequency, with the frequency dependence strongest on the approach to \tc.  Below a temperature $T^\ast = 2.3$~K, $1/\lambda^2(T)$ crosses over to a quadratic temperature dependence, the characteristic behaviour of $d$-wave superconductivity in the presence of disorder.  Inset: $1/\lambda^2(T)$ at 2.50~GHz, plotted vs.\ $T^2$.  } 
\label{fig:rhos}
\end{figure}

The frequency-dependent superfluid density, $1/\lambda^2$, is obtained from the imaginary part of the microwave conductivity according to:
\begin{equation}
\frac{1}{\lambda^2(\omega,T)} \equiv \omega \mu_0 \sigma_2(\omega,T)\;.
\end{equation}
Superfluid density is plotted in Fig.~\ref{fig:rhos} for the 17 frequencies. The most striking feature is the strong linear temperature dependence extending from 2 to 14~K.  While $\Delta \rho_s(T) \propto T$ is the hallmark of line nodes in the energy gap, the observed behaviour extends over a wider range than expected.   A similarly strong temperature dependence is seen in near-optimally doped \tbcod,\cite{Broun:1997p387} and underdoped \ybco{6.333},\cite{Broun:2007p49} but not in optimally doped \ybco{y}, \cite{hardy93,Harris:2006p388} or \bscco{8+\delta}.\cite{Lee:1996p382}  In overdoped \tbcod\ at higher temperatures, $1/\lambda^2(T)$ has upward curvature: the behaviour in the vicinity of 20~K is almost certainly not intrinsic and, as with the rounding of $R_s(T)$, is likely associated with spatial variation of \tc\ due to chemical inhomogeneity.  In fact, the $1/\lambda^2(T)$ data allow us to place experimental bounds on the variation of $T_c$ in the sample, confining it to the range 20~to~25~K.  This is at odds with the model of microscopic inhomogeneities\cite{Geshkenbein:1998p3010} put forward to explain the anomalous behaviour of the resistive critical field,\cite{MACKENZIE:1993p197} and it would therefore be very interesting to repeat the magnetotransport measurements on a similar sample to that measured here, to see if improvement in the homogeneity of dopant oxygen atoms has changed the form of $B_\rho(T)$.  Returning to the superfluid density data, the frequency variation of $1/\lambda^2$ is also substantial in the vicinity of the superconducting transition, and this is in keeping with the strong frequency variation of $\sigma_1$ at similar temperatures: the microwave conductivity is a causal response function, and its real and imaginary parts must be related by Kramers--Kr\"onig relations.  At the lowest temperatures, $1/\lambda^2(T)$ crosses over to a quadratic temperature dependence, as shown in the inset of Fig.~\ref{fig:rhos}.  This behaviour is consistent with the pair-breaking effects of disorder in a $d$-wave superconductor and will be discussed further below.

\section{Discussion}
\label{Sec:discussion}

The recent availability of high quality normal-state data for overdoped \tbcod\ enables a quantitative analysis of the superconducting impurity physics to be carried out.  Of particular interest are de Haas--van Alphen  (dHvA) measurements made on $T_c = 26$~K overdoped \tbcod.\cite{Bangura:2010p1675}  For magnetic field applied along the $c$-direction, the dHvA measurements find a fundamental frequency $F_0 = 17.63$~kT, corresponding to an extremal Fermi-surface cross-sectional area $\pi k_F^2 = \mathcal{A} = 2 \pi e F_0/\hbar$, yielding an average Fermi wave vector $k_F = 7.3 \times 10^9$~m$^{-1}$.  For $T_c = 26$~K \tbcod, the hole-doping inferred from the dHvA measurements is $p = 0.270$, corresponding to a carrier concentration \mbox{$n = 2 (1 + p)/a^2 c = 7.35 \times 10^{27}$~m$^{-3}$}, where $a = 3.86$~\AA\ and $c = 23.2$~\AA\ are the in-plane and inter-plane lattice parameters, respectively, and the prefactor of 2 takes into account that there are 2 CuO$_2$ planes in the body-centred tetragonal unit cell. The second important piece of data from the dHvA study is the quasiparticle mass, $m^\ast$, obtaining by fitting Lifshitz--Kosevich theory to the temperature damping of the oscillation amplitude.  For the $T_c = 26$~K sample, $m^\ast = 5.0 \pm 0.3~m_e$ is obtained.

The in-plane DC resistivity of the sample used in our microwave study is 10.2~$\mu\Omega$cm at $T = 27.5$~K. From magnetotransport measurements on similar samples,\cite{Proust:P2lqZi4f} we know that the resistivity should decrease by a further 3.6~$\mu\Omega$cm on cooling to $T=0$, implying a residual resistivity $\rho_0 \approx 6.6~\mu\Omega$cm for our sample.
The elastic contribution to transport relaxation time is therefore \mbox{$\tau_n = m^\ast/n e^2 \rho_0 = 3.7 \times 10^{-13}$~s}.  The normal-state scattering-rate parameter relevant for the SCTMA theory is \mbox{$\Gamma_n = (2 \tau_n)^{-1} = 1.37 \times 10^{12}$~s$^{-1}$}.  In temperature units, \mbox{$\hbar \Gamma_n/k_B = 10.4$~K}.  From the dHvA measurements we estimate a Fermi velocity \mbox{$v_F = \hbar k_F/m^\ast = 1.70 \times 10^5$~m/s}.  The elastic mean free path from transport is then \mbox{$\ell = v_F \tau_n = 620$~\AA}, with $k_F \ell = 450$.  This can be compared to the single-particle elastic mean free path estimated from a Dingle analysis of the dHvA measurements, \mbox{$\ell_0 = 360$~\AA}.  Transport mean free path $\ell$ probes momentum relaxation, and its significant enhancement over the single-particle mean free path $\ell_0$ indicates that small angle scattering processes are important, which is consistent with the cation disorder in \tbcod\ being located away from the CuO$_2$ planes.  The single-particle scattering rate is, however, more appropriate for comparisons with SCTMA theory, where the dominant physics is that of impurity pair breaking.  Using the Dingle mean free path we instead obtain \mbox{$\Gamma_n = (2 \tau_n)^{-1} = 2.35 \times 10^{12}$~s$^{-1}$},  \mbox{$\hbar \Gamma_n/k_B = 18$~K}, and $\hbar \Gamma_n/k_B T_c \approx 0.7~\mathrm{to}~1.0$, depending on the value of $T_c$ used in the normalization.  We note, too, that the Dingle mean free path provides a lower bound on $\Gamma_n$, since the quantum oscillation measurements self-select the best regions of the best samples.

\begin{figure}[t]
\centering
\includegraphics[width = \figwidthtwo \columnwidth]{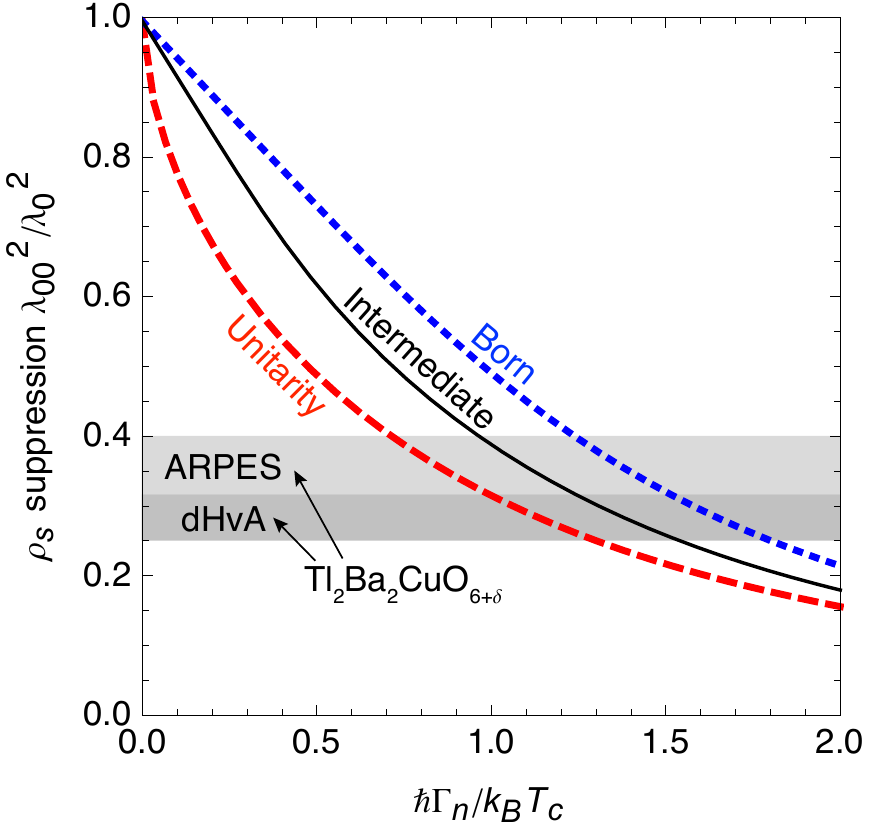}
\caption{(Color online)  The suppression of zero-temperature superfluid density by strong-, intermediate- and weak-scattering disorder.  The suppression fraction is $\lambda_{00}^2/\lambda_0^2$, where $\lambda_{00}$ is the zero-temperature penetration depth with no disorder, and $\lambda_0$ is the zero-temperature penetration depth in the presence of disorder.  Data are shown for strong ($c=0$, long dashes), intermediate ($c=1$, solid line) and weak ($c \gg 1$, short dashes) scattering regimes, as a function of the corresponding normal-state scattering rate $\Gamma_n$.  The degree of suppression relevant to overdoped \tbcod\ is indicated by the shaded bands, one based on an estimate of $\lambda_{00}$ from the ARPES energy dispersion, the other from dHvA data.} 
\label{fig:suppression}
\end{figure} 

\begin{figure}[t]
\centering
\includegraphics[width = \figwidthtwo \columnwidth]{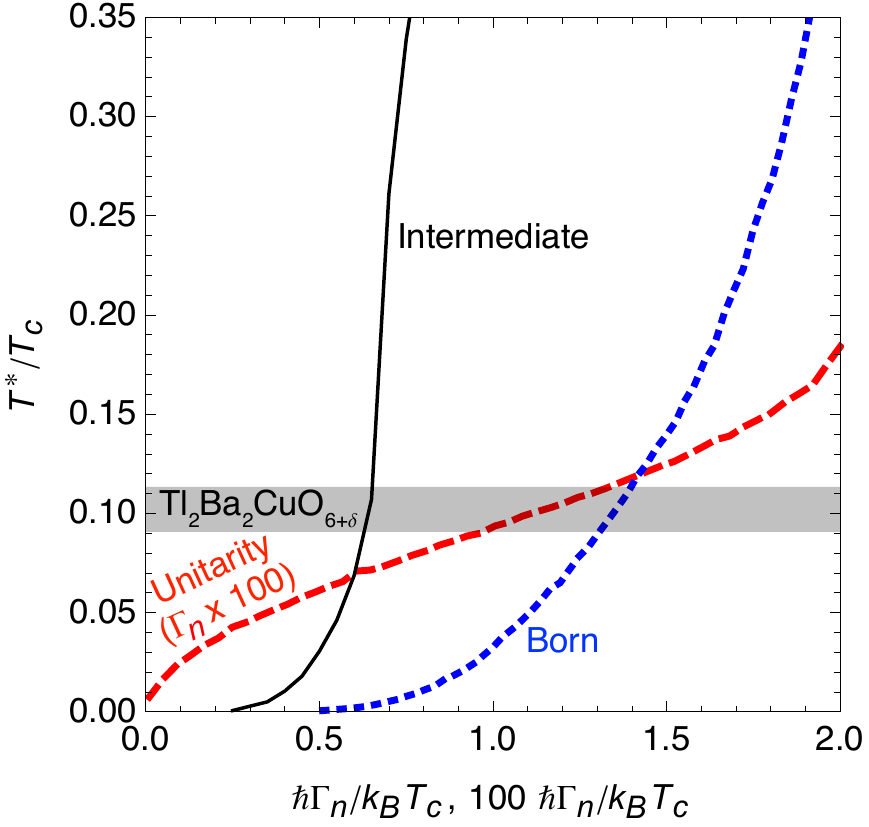}
\caption{(Color online) Dependence of the superfluid-density disorder-crossover temperature, $T^\ast$, on the normal-state scattering rate, $\Gamma_n$, for strong ($c=0$, long dashes), intermediate ($c=1$, solid line) and weak ($c \gg 1$, short dashes) scattering disorder. Note that the horizontal scale has been expanded ($\Gamma_n$ has been multiplied by a factor of 100) for the strong-scattering, unitarity-limit curve.  The parameter regime relevant to overdoped \tbcod\ ($T^\ast \approx 0.1~T_c$) is indicated by the shaded band.  Only for weak-to-intermediate-strength scattering does the implied value of $\Gamma_n$ agree with that obtained from normal-state transport.}
\label{fig:tstar}
\end{figure} 

The availability of detailed normal-state data on \tbcod\ allows us to estimate the degree to which the zero-temperature superfluid density has been suppressed from its ideal, pure value.  In the pure case, the zero-temperature penetration depth is given by \mbox{$1/\lambda_{00}^2 \approx \mu_0 ne^2/m^\ast$}.  Using values for $n$ and $m^\ast$ from the dHvA experiments, we obtain $\lambda_{00} \approx 1390\pm3$\%~\AA\ for \mbox{$T_c = 26$~K} \tbcod.  An alternative estimate can be made using the ARPES energy dispersion --- this has been carried out in Appendix~B for $T_c \approx 30$~K material, for which we obtain  $\lambda_{00} = 1560$~\AA.  

The measured zero-temperature penetration depth in the real material is $\lambda_0 = 2610 \pm 5$\%~\AA, implying that superfluid density has been suppressed to $\lambda_{00}^2/\lambda_0^2 \approx 25$--32\% of its pure value based on the dHvA data, or 32--40\% based on the ARPES dispersion.  Note that while \tbcod\ technically satisfies the requirements of a clean-limit superconductor \mbox{($\xi_0 < \ell$ or $\hbar \Gamma_N < \Delta$)} it is nevertheless in a regime in which impurity scattering is expected to cause substantial suppression of superfluid density.  To illustrate this, Fig.~\ref{fig:suppression} shows theoretical curves for $\lambda_{00}^2/\lambda_0^2$ in a $d$-wave superconductor, for point-like impurities of strong, weak and intermediate scattering strengths.  Note that the curves are plotted as a function of the normal-state scattering rate $\Gamma_n$, and a given value of $\Gamma_n$ can be either be achieved with a low density of strong-scattering impurities, or a high density of weak-scattering impurities.  (The curves have been evaluated using results from SCTMA theory for point-like disorder, with details given in Appendix~A.  Scattering strength is parameterized by $c$, the cotangent of the scattering phase shift.)  The experimentally relevant ranges for \tbcod\ are shown as shaded bands, and are close to expectations for impurities acting in the intermediate-strength scattering regime.  It therefore seems unnecessary to invoke models of inhomogeneous superconductivity to explain the suppression of superfluid density on the overdoped side of the phase diagram\cite{Uemura:1993gs,2001SSCom.120..347U} --- this will naturally occur as the superconducting gap closes and the material gets pushed toward the dirty limit.

For a pure $d$-wave superconductor, the expected behaviour of the superfluid density is a linear temperature dependence at low temperatures, due to the presence of line nodes in the energy gap.  Disorder gives rise to pair breaking, causing a crossover to quadratic temperature dependence at low temperatures.  This can be modelled with the cross-over formula\cite{HIRSCHFELD:1993tf}
\begin{equation}
\rho_s(T) = \rho_0 - A \frac{T^2}{T + 2 T^\ast}\;,
\label{Eq:crossover}
\end{equation}
where the cross-over temperature $T^\ast$ has been defined to be the point at which linear and quadratic regimes have the same temperature slope.  (This definition of $T^\ast$ differs by a factor of 2 from that used in SCTMA studies,\cite{HIRSCHFELD:1993tf} but is a natural choice for comparing with experiment.)  From our measurements on \tbcod, we find $T^\ast = 2.3$~K, and therefore $T^\ast/T_c \approx 0.1$.  In order to examine how cross-over temperature is influenced by impurity physics, we have evaluated $\rho_s(T)$ in the SCTMA theory for disorder of different scattering strengths and impurity concentrations, and then have fit $\rho_s(T)$ to Eq.~(\ref{Eq:crossover}) to obtain $T^\ast$.  Results are plotted in Fig.~\ref{fig:tstar} for three different scattering regimes, as a function of $\hbar\Gamma_n/k_B T_c$.  Unlike the suppression of superfluid density (shown in Fig.~\ref{fig:suppression}), whose dependence on $\Gamma_n$ is not particularly sensitive to the type of impurity, the disorder cross-over temperature shows marked variation, and is therefore very useful for identifying the strength of scattering, if $\Gamma_n$ is independently known.  The range of scattering rate parameter relevant to \tbcod\ is $\hbar \Gamma_n/k_B T_c \approx 0.7~\mathrm{to}~1.0$. We see in Fig.~\ref{fig:tstar}  that this intersects with $T^\ast/T_c \approx 0.1$ in the weak-scattering regime, very far from the unitarity-limit curve, which has been expanded horizontally by a factor of 100 to be visible on the graph.

\begin{figure}[t]
\centering
\includegraphics[width = \figwidthtwo \columnwidth]{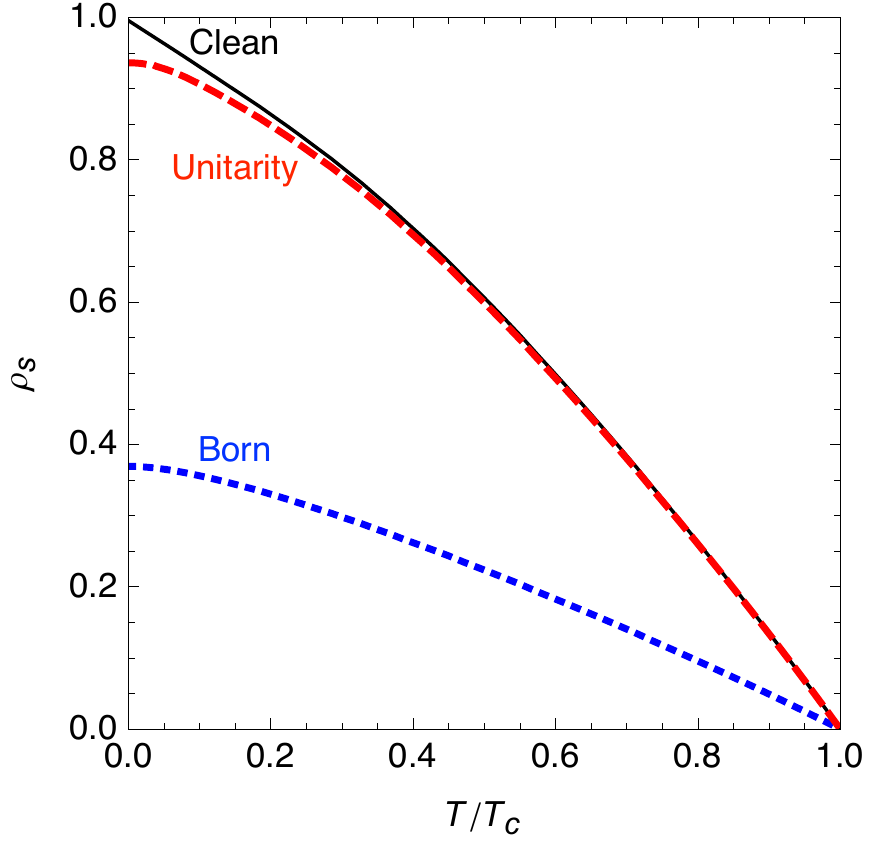}
\caption{(Color online) Effect of strong- and weak-scattering disorder on superfluid density, $\rho_s$, in a $d$-wave superconductor, from SCTMA theory.  In the absence of disorder, $\rho_s$ has linear $T$ dependence at low $T$ (clean limit, solid curve).  The pair-breaking effects of strong scattering disorder (unitarity limit, long dashes) are confined to low temperatures, with $\Delta \rho_s(T) \propto T^2$ below a cross-over temperature $T^\ast$.  Weak-scattering disorder (Born limit, short dashes) can drive a similar crossover to quadratic $T$ dependence, but only at such a high density of scatterers that the $\rho_s(T)$ is suppressed from its clean-limit form over the entire temperature range.  For both unitarity ($c = 0,~\Gamma = 0.01~T_c$) and Born-limit curves ($c = 30,~\Gamma = 1200~T_c$), the level of disorder has been adjusted so that $T^\ast \approx 0.1~T_c$, the regime relevant to overdoped \tbcod.} 
\label{fig:rhosSCTMA}
\end{figure} 

\begin{figure}[t]
\centering
\includegraphics[width = \figwidthtwo \columnwidth]{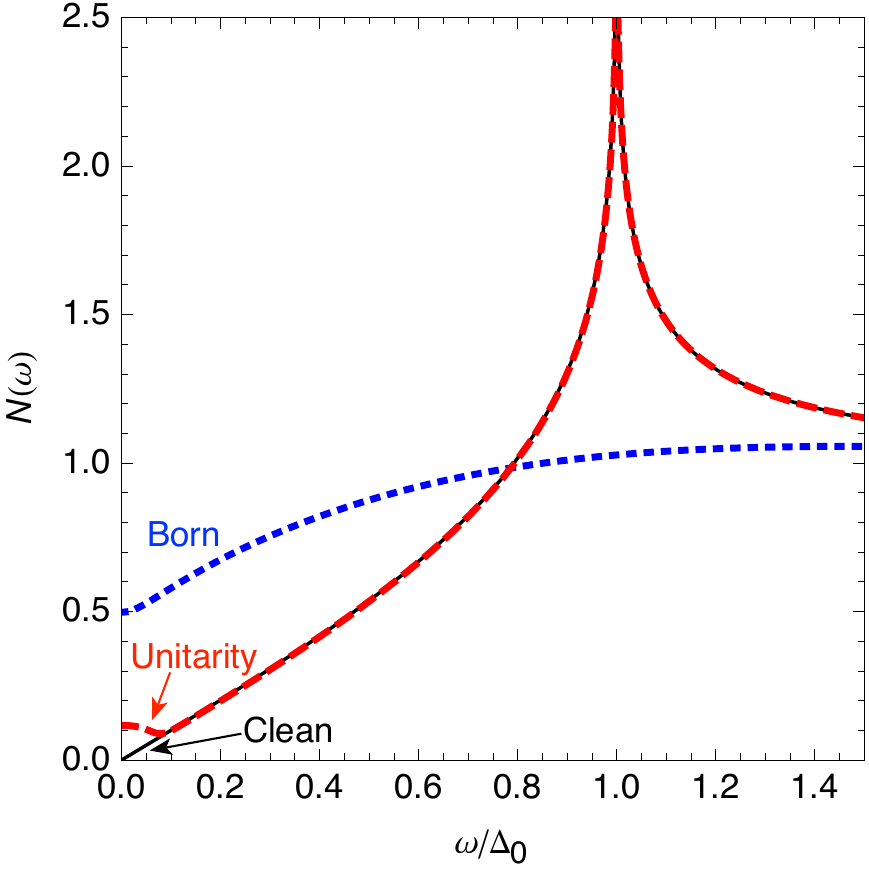}
\caption{(Color online) The normalized density of states, $N(\omega)$, of a $d$-wave superconductor at $T = 0$, in the presence of weak- and strong-scattering disorder.  SCTMA results have been evaluated for clean limit (solid line), unitarity limit (long dashes) and Born limit (short dashes) cases, for the same set of scattering parameters as in Fig.~\ref{fig:rhosSCTMA}.  The pair-breaking effects of strong-scattering disorder are confined to a narrow range of energies near $\omega = 0$.  For the Born limit, which is most appropriate to \tbcod, the density of states bears little resemblance to that of a clean-limit superconductor, with no coherence peak at $\omega = \Delta_0$.} 
\label{fig:DOS}
\end{figure} 

To provide another perspective on the impurity physics underlying $\rho_s(T)$, Fig.~\ref{fig:rhosSCTMA} compares theoretical plots of $\rho_s(T)$ for three cases: the pure $d$-wave superconductor; the same superconductor with strong-scattering impurities and $T^\ast/T_c = 0.1$; and, keeping $T^\ast/T_c = 0.1$, a $d$-wave superconductor with weak-scattering impurities.  In the strong-scattering case, the effect of the disorder is localized in energy, as can be seen in the corresponding plot of density of states, $N(\omega)$, in Fig.~\ref{fig:DOS}.  This localized pair breaking drives the crossover to quadratic behaviour with little effect on the absolute superfluid density.  By contrast, in the weak-scattering case, the absolute magnitude of the superfluid density is strongly suppressed, in line with what is observed in \tbcod.  The corresponding density of states in Fig.~\ref{fig:DOS} is barely recognizable as a superconductor, with a large residual density of states, $N(0)$, and the notable absence of a coherence peak near $\omega = \Delta_0$.  Surprising, the overall shape of $\rho_s(T)$ is very similar in weak- and strong-scattering cases, pointing to the difficulty of identifying impurity parameters from qualitative features in the superfluid density and penetration depth: absolute numbers are required, as is a good knowledge of the normal state.

In a $d$-wave superconductor, scattering of quasiparticles by impurities is counterbalanced by the pair-breaking effects of disorder, leading to bare transport coefficients that are universal in the zero-temperature, zero-frequency limit.\cite{Lee:1993ky}  This is well-established in the case of the residual thermal conductivity, $\kappa_0$, which has been shown to obey\cite{Norman:1996gx,Graf:1996ix,Taillefer:1997it,Durst:2000p963,Chiao:2000p2655,Shakeripour:2009is}
\begin{equation}
\frac{\kappa_0}{T} = \frac{k_B^2}{3 \hbar} \frac{1}{d} \left(\frac{v_F}{v_\Delta} + \frac{v_\Delta}{v_F} \right) \approx \frac{k_B^2}{3 \hbar} \frac{1}{d}\frac{v_F}{v_\Delta}\;.
\label{Eq:universal_kappa}
\end{equation}
Here $d$ is the average spacing between CuO$_2$ planes and $v_\Delta  = (\hbar k_F)^{-1} \D \Delta/\D \phi$ is the gap velocity, determined by the energy dispersion parallel to the Fermi surface, near the node.  For heavily overdoped cuprates, $v_\Delta \sim 0.01~v_F$.\cite{Hawthorn:2007hq}

The situation for the electrical conductivity is more complicated.  To first approximation, the Wiedemann--Franz law applies, giving a bare universal conductivity\cite{Lee:1993ky} 
\begin{equation}
\sigma_{00} = \frac{3}{\pi^2} \frac{e^2}{k_B^2} \frac{\kappa_0}{T} = \frac{e^2}{\pi^2 \hbar^2} \frac{1}{d} \frac{v_F}{v_\Delta}\;.
\end{equation}
However, Durst and Lee have shown\cite{Durst:2000p963} that the electrical conductivity is modified by a vertex-correction parameter, $\beta_\mathrm{VC}$, due to small-angle-scattering effects, and a Fermi-liquid correction, ${\alpha_\mathrm{FL}^s}$:
\begin{equation}
\sigma_0 = \sigma_{00} \beta_\mathrm{VC} {\alpha_\mathrm{FL}^s}^2\;.
\end{equation}
$\sigma_{00}$ can be obtained from estimates of $v_\Delta$ but, since the best data on $v_\Delta$ typically come from thermal conductivity, it is simplest to start directly with measurements of residual thermal conductivity and apply the Wiedemann--Franz law.  For $T_c \approx 25$~K \tbcod, thermal conductivity measurements give \mbox{$\kappa_0/T \approx 0.34~\mathrm{to}~0.48$~mW~K$^{-2}$cm$^{-1}$} in the $T \to 0$ limit,\cite{Hawthorn:2007hq} corresponding to $\sigma_{00} \approx 1.4~\mathrm{to}~2.0 \times 10^6~\Omega^{-1}\mathrm{m}^{-1}$.  This range of $\sigma_{00}$ is indicated on the conductivity plots earlier in the paper.  The measured residual conductivity in \tbcod\ is approximately 6 times higher than $\sigma_{00}$.  It is difficult to separate vertex-correction and Fermi-liquid contributions to this enhancement, as the temperature dependence of the superfluid density (the other quantity in which Fermi-liquid corrections occur) is strongly affected by disorder in the Born limit, as shown in Fig.~\ref{fig:rhosSCTMA}.  Nevertheless, it is likely that vertex corrections play an important role: the enhancement of transport mean free path over single-particle mean free path in the normal state indicates the importance of small-angle scattering above $T_c$; the dominant cation disorder in \tbcod\ is located away from the CuO$_2$ planes, softening the impurity potentials; and the nodal structure of the $d$-wave energy spectrum prevents superconducting quasiparticles from diffusing around the Fermi surface by a succession of small-momentum-transfer scattering events.\cite{Durst:2000p963} A shortcoming of the SCTMA approach we compare to in this paper is the assumption of point-like disorder --- the model does not allow for extended defects, which are very relevant for the type of off-plane disorder present in \tbcod.  The extension of the SCTMA formalism\cite{Durst:2000p963,Nunner:2005p654} that has been developed to capture the momentum-dependent scattering arising from extended defects has so far been applied to the calculation of transport coefficients such as electrical and thermal conductivity.  It would be very useful if this approach could be extended to study the correlation between superfluid density suppression, disorder cross-over temperature and normal-state scattering rate.  That is, it is important to know whether the plots in Figs.~\ref{fig:suppression} and \ref{fig:tstar}, which were calculated for point-like impurities, are modified by the inclusion of extended defects. In the case of \tbcod, for which the nature and concentration of the dominant cation and dopant disorder are well known, the relevant impurity parameters can likely be constrained tightly by first principles calculations.  This is turn should settle the issue as to whether microscopic phase separation and inhomogeneity are necessary to understand the strongly overdoped cuprates.

Finally, we return to the issue of superfluid suppression and consider the uncondensed spectral weight, which must appear in the finite-frequency conductivity $\sigma_1(\omega)$.  A sum-rule argument gives residual conductivity spectral weight
 \begin{equation}
I_0 = \int_{0_+}^{\omega_c} \!\!\!\!\!\!\sigma_1(\omega,T \to 0) \D \omega = \frac{\pi}{2} \frac{n_n e^2}{m^\ast} = \frac{\pi}{2} \frac{1}{\mu_0} \left(\frac{1}{\lambda_{00}^2} - \frac{1}{\lambda_0^2} \right)\;,
\end{equation}
where $n_n$ is the effective normal-fluid density and $\omega_c$ is a frequency cut-off chosen to include contributions from conduction electrons while avoiding interband transitions, etc.  $I_0$ can be used to estimate the width of the quasiparticle conductivity spectrum and therefore the average quasiparticle relaxation rate at low temperatures.  Assuming, for simplicity, a Drude form for the quasiparticle conductivity, \mbox{$\sigma_1(\omega) = \sigma_0/(1 + \omega^2/\Gamma_0^2)$}, we have
\begin{equation}
I_0 = \int_0^\infty \frac{\sigma_0}{1 + \omega^2/\Gamma_0^2} \D \omega = \tfrac{\pi}{2} \sigma_0 \Gamma_0\;,
\end{equation}
leading to $\Gamma_0 = (1/\lambda_{00}^2 - 1/\lambda_0^2)/\mu_0 \sigma_0$.  From our data, for which $\sigma_0 \approx 10^7~\Omega^{-1}\mathrm{m}^{-1}$,
we obtain $\Gamma_0/2\pi \approx 500$~GHz.  This is very broad, extending well beyond our measurement range and confirming that the narrow, low-frequency component observed below $T_c$ in $\sigma(\omega)$ cannot originate from quasiparticle relaxation.

\section{Conclusions}

We have measured the microwave conductivity and superfluid density of $T_c = 25$~K \tbcod, extending the range of this important probe of pairing symmetry and quasiparticle charge dynamics deep into the overdoped regime.  The observed linear temperature dependence of the superfluid density indicates a pairing state with line nodes, as expected for a $d$-wave superconductor.  The relative magnitude of the linear term in $\rho_s(T)$ is large, and extends over most of the superconducting temperature range: this is unusual, but is similar to what has been observed in \tbcod\ near optimal doping.\cite{Broun:1997p387} The disorder cross-over temperature observed in $\rho_s(T)$ is relatively small ($T^\ast \approx 0.1~T_c$), giving the superficial impression of a clean-limit $d$-wave superconductor.  But the normal-state scattering rate and the suppression of superfluid density are substantial, indicating the presence of a high density of weak-to-intermediate-strength scatterers. Small-angle scattering effects are readily apparent in the normal-state transport and in the enhancement of the residual conductivity over the bare universal value for a $d$-wave superconductor.  Taken together, these observations are consistent with the known cation disorder in \tbcod\ --- an excess of Cu that substitutes onto Tl sites  well separated from the CuO$_2$ planes.  

A coherent picture is emerging from overdoped \tbcod, in which a concentration of measurements on crystals of comparable quality and hole-doping is allowing different measurements to be brought together to extract more detailed information than would otherwise be possible.  In the current context, the availability of detailed Fermi-surface information allows a critical comparison of the microwave data with SCTMA theory, which forms the standard model of Fermi-liquid-based superconductivity in the presence of disorder.  

\section{Acknowledgements}

The authors thank W.~A.~Atkinson for useful discussions and M.~L.~W.~Thewalt for supplying the ultrahigh-purity silicon used in the sample holders.  Research support for the experiments was provided by the Natural Science and Engineering Research Council of Canada (NSERC) and the Canadian Foundation for Innovation.  Research support for sample preparation was provided by NSERC and the Canadian Institute for Advanced Research.\\

\appendix

\section{DIRTY D-WAVE SUPERCONDUCTIVITY}\label{Sec:AppendixA}

The self-consistent $t$-matrix approximation (SCTMA) provides a powerful framework for describing the pair-breaking effects of disorder in a \dwave\ superconductor.\cite{NAM:1967p640,PETHICK:1986p562,HIRSCHFELD:1988p611,PROHAMMER:1991p557,Schachinger:2003p635,HIRSCHFELD:1993tf,HIRSCHFELD:1993p567,BORKOWSKI:1994p647,Joynt:1997p601,Hussey:2002p649,Nunner:2005p654,Balatsky:2006p607}  In this appendix we describe the evaluation of SCTMA results for the normalized quasiparticle density of states, $N(\omega)$, and superfluid density, $1/\lambda^2(T)$, in the presence of strong- and weak-scattering disorder.  In the SCTMA, impurities are treated as isotropic point scatterers.  In a \dwave\ superconductor, disorder renormalizes the quasiparticle energies $\omega$ to $\tilde\omega(\omega)$, according to 
\begin{equation}
\omega \to \tilde\omega(\omega)  = \omega + \I \Gamma \frac{N(\omega,T)}{c^2 + N^2(\omega,T)}\;,
\label{renorm1}
\end{equation}
where $c$ is the cotangent of the \swave\ scattering phase shift, $\Gamma = n_i n/\pi D(\epsilon_F)$, $n_i$ is the impurity concentration, $n$ is the conduction electron density, and $D(\epsilon_F)$ is the density of states at the Fermi level.\cite{HIRSCHFELD:1993tf}    Resonant (unitarity-limit) scattering corresponds to $c = 0$; weak (Born-limit) scattering to $c \gg 1$.

For the evaluations, we make the simplifying assumptions of an isotropic, circular, 2D Fermi surface, and an energy gap $\Delta_\mathbf{k}$ that takes the form of the simplest cylindrical harmonic with $d_{x^2 - y^2}$ symmetry,
\begin{equation}
\Delta_\mathbf{k} = \Delta_0(T) \cos 2 \phi\;.
\end{equation}
Here $\phi$ measures angle from the Cu--O bond direction. For the temperature dependence of the energy gap, $\Delta_0(T)$, we use the interpolation formula\cite{Prozorov:2006wr}
\begin{equation}
\Delta_0(T) = \Delta_0(0) \tanh\left(\frac{\pi k_B T_c}{\Delta_0(0)} \sqrt{a\left(\frac{T_c}{T} - 1\right)}\right)\;.
\end{equation}
The appropriate parameters for a weak-coupling \dwave\ superconductor are $\Delta_0(0) = 2.14~k_B T_c$ and $a = \frac{4}{3}$.\cite{Prozorov:2006wr}  

The quasiparticle density of states is
\begin{equation}
N(\omega,T)  = \left\langle \frac{\tilde\omega}{\sqrt{\tilde\omega^2 - \Delta_0^2(T) \cos^2 2 \phi}}\right\rangle_{\!\!\phi} = \frac{2}{\pi} K\!\left(\frac{\Delta_0^2(T)}{\tilde\omega^2} \right)\;,\label{dos}
\end{equation}
where $\langle ... \rangle_{\phi}$ denotes an angle average around the cylindrical Fermi surface and $K(x)$ is the complete elliptic integral of the first kind.  The branch of the square root in Eq.~(\ref{renorm1}) is chosen to give $\tilde \omega$ positive imaginary part.  Although not shown explicitly, $\tilde \omega$ has temperature dependence, arising from the temperature dependence of $\Delta_0$, and is re-evaluated  at each temperature in the plots in Figs.~\ref{fig:tstar} and \ref{fig:rhosSCTMA}.

Normalized superfluid density is given by\cite{HIRSCHFELD:1993tf}
\begin{equation}
\frac{\lambda^2_{00}}{\lambda^2(T)}\!\! =  \!\!\tfrac{1}{2}\!\!\int_{-\infty}^\infty \!\!\!\!\!\!\D \omega \tanh\!\left(\frac{\beta \omega}{2}\right)\!\mathrm{Re} \left\langle\!\frac{\Delta^2(\phi,T) }{\big(\tilde\omega^2 - \Delta^2(\phi,T)\big)^\frac{3}{2}}\!\right\rangle_{\!\!\phi}\;,
\label{dirtylambda}
\end{equation}
where $\beta = 1/k_B T$.  The thermal average is most efficiently evaluated using the Matsubara sum\cite{PROHAMMER:1991p557}
\begin{equation}
\frac{\lambda^2_{00}}{\lambda^2(T)} =  \frac{2 \pi}{\beta} \sum_{\;\;\I \omega_n} \mathrm{Re} \left\langle\!\frac{\Delta^2(\phi,T) }{\big(\tilde\omega^2(\I \omega_n) - \Delta^2(\phi,T)\big)^\frac{3}{2}}\!\right\rangle_{\!\!\phi}\;,
\label{Eq:Matsubara}
\end{equation}
where the Matsubara frequencies, $\omega_n = (2n + 1)\pi/\beta$, range over positive integers $n$.  The density of states factor in Eq.~(\ref{Eq:Matsubara}) is  
\begin{equation}
\begin{split}
\left\langle \frac{\Delta_0^2(T) \cos^2 2 \phi}{\big(\tilde\omega^2 - \Delta_0^2(T) \cos^2 2 \phi\big)^\frac{3}{2}}\right\rangle_{\!\!\phi}= \qquad\qquad \\
 \frac{2}{\pi \tilde\omega}\left[K\left(\frac{\Delta_0^2(T)}{\tilde\omega^2}\right) + \frac{\tilde\omega^2 }{\Delta_0^2(T) - \tilde\omega^2}\;E\left(\frac{\Delta_0^2(T)}{\tilde\omega^2}\right)\right],
\end{split}
\end{equation}
where $E(x)$ is the complete elliptic integral of the second kind.

\section{ARPES DISPERSION AND ZERO TEMPERATURE SUPERFLUID DENSITY}\label{Sec:AppendixB}

The usual expression for zero-temperature superfluid density,
\begin{equation}
\frac{1}{\lambda_{00}^2} = \frac{\mu_0 n_s e^2}{m^\ast}\;,
\label{Eq:superfluid}
\end{equation}
is exact only in the isotropic case of a cylindrical or spherical Fermi surface.  It is nevertheless convenient, as it draws on parameters that can be measured directly using quantum oscillations, namely the Fermi surface volume and the quasiparticle cyclotron mass.  In this Appendix, we provide an alternative estimate of the zero-temperature penetration depth, based on a tight-binding parameterization of the ARPES energy dispersion in $T_c \approx 30$~K \tbcod:\cite{Plate:2005p2658}
\begin{equation}
\begin{split}
\epsilon_{\mathbf{k},\parallel} & \approx \mu + \tfrac{t_1}{2} \big( \cos (k_x a) + \cos(k_y a) \big) + t_2 \cos(k_x a) \cos(k_y a)\\
&+  \tfrac{t_3}{2} \big( \cos (2 k_x a) + \cos(2 k_y a) \big) +  \tfrac{t_4}{2} \big( \cos(2 k_x a) \cos( k_y a) \\
& +  \cos(k_x a) \cos( 2 k_y a)\big) +  t_5 \cos(2 k_x a) \cos( 2 k_y a),
\end{split}
\label{Eq:dispersion}
\end{equation}
with $\mu = 243.8$, $t_1 = -725$,  $t_2 = 302$, $t_3 = 15.9$, \mbox{$t_4 = -80.5$} and $t_5 = 3.4$~meV.  This results in a single, large Fermi surface, closing around $\mathbf{k} = (\pi/a,\pi/a)$ and containing $1 + p = 1.27$~holes per planar copper.

The zero-temperature penetration depth is closely related to the plasma frequency, $\omega_p$, and is given in the general, anisotropic case by\cite{CHANDRASEKHAR:1993p329}
\begin{equation}
\frac{1}{\lambda_{00}^2} \equiv \mu_0 \epsilon_0 \omega_p^2 = 2 \mu_0 e^2 \!\!\!\int \frac{\D^3\mathbf{k}}{(2 \pi)^3} \delta(\epsilon_\mathbf{k} - \epsilon_F) v_{\mathbf{k},x}^2.
\end{equation}
Here, the prefactor of 2 is due to spin degeneracy, and the $x$-component of velocity $v_\mathbf{k}$ is taken in order to obtain in-plane penetration depth.  For an isotropic energy dispersion in 2D or 3D, $\epsilon_\mathbf{k} = \hbar^2 k^2/2 m^\ast$, it is straightforward to show that the usual expression, Eq.~(\ref{Eq:superfluid}), results.  Using the energy dispersion appropriate to overdoped \tbcod, Eq.~(\ref{Eq:dispersion}) and assuming a weak modulation along the $c$ direction, we obtain $\lambda_{00} = 1560$~\AA.  Note that this result only includes renormalization effects present in the bare ARPES energy dispersion and not, for example, additional effects arising from Fermi-liquid interactions.  For completeness, we note that the tight-binding ARPES dispersion gives the same carrier density as quantum oscillation measurements on $T_c \approx 26$~K \tbcod, and therefore the same average Fermi wavevector, $k_F = 7.3 \times 10^9$~m$^{-1}$.  The angle-averaged Fermi velocity, $v_F = 1.3 \times 10^5$~m/s, is lower than the dHvA value, consistent with the fact that the ARPES dispersion implies a slightly larger cyclotron mass, 
\begin{equation}
m_c = \frac{\hbar}{2 \pi} \oint_\mathrm{FS} \frac{\D \mathbf{k}}{| v_\mathbf{k}|} = 6.7~m_e\;.
\end{equation}


%

\end{document}